\documentclass[english,a4paper,preprint]{aastex}
\usepackage[T1]{fontenc}
\usepackage[latin9]{inputenc}
\usepackage{graphicx}
\usepackage{amssymb}

\makeatletter

\providecommand{\tabularnewline}{\\}

\slugcomment{Submitted to Icarus.}

\shorttitle{V-type asteroids in the middle Main Belt}
\shortauthors{Roig et al.}

\usepackage{babel}
\makeatother

\begin{document}

\title{V-type asteroids in the middle Main Belt\altaffilmark{1}}

\author{F. Roig}

\affil{Observatório Nacional, Rio de Janeiro, Brazil}

\author{D. Nesvorný\altaffilmark{2}}

\affil{Southwest Research Institute, Boulder, CO, USA}

\author{R. Gil-Hutton}

\affil{Complejo Astronómico El Leoncito (CASLEO) and Universidad Nacional
de San Juan, Argentina}

\and{}

\author{D. Lazzaro}

\affil{Observatório Nacional, Rio de Janeiro, Brazil}

\affil{}

\altaffiltext{1}{Based on observations obtained at the Gemini Observatory,
which is operated by the Association of Universities for Research
in Astronomy, Inc., under a cooperative agreement with the NSF on
behalf of the Gemini partnership: the National Science Foundation
(USA), the Particle Physics and Astronomy Research Council (UK), the
National Research Council (Canada), CONICYT (Chile), the Australian
Research Council (Australia), CNPq (Brasil) and CONICET (Argentina)
- Program ID: GS-2006A-Q51.}

\altaffiltext{2}{Visiting Scientist, Observatório Nacional, Rio
de Janeiro, Brazil}

\begin{abstract}
V-type asteroids are bodies whose surfaces are constituted of basalt.
In the Main Asteroid Belt, most of these asteroids are assumed to
come from the basaltic crust of asteroid (4) Vesta. This idea is mainly
supported by (i) the fact that almost all the known V-type asteroids
are in the same region of the belt as (4) Vesta, i.e. the inner belt
(semi-major axis $2.1<a<2.5$ AU), (ii) the existence of a dynamical
asteroid family associated to (4) Vesta, and (iii) the observational
evidence of at least one large craterization event on Vesta's surface.
One V-type asteroid that is difficult to fit in this scenario is (1459)
Magnya, located in the outer asteroid belt, i.e. too far away from
(4) Vesta as to have a real possibility of coming from it.

The recent discovery of the first V-type asteroid in the middle belt
($2.5<a<2.8$ AU), (21238) 1995WV7 (Binzel et al. 2006, DPS 38, \#71.06;
Hammergren et al. 2006, astro-ph/0609420), located at $\sim2.54$
AU, raises the question of whether it came from (4) Vesta or not.
In this paper, we present spectroscopic observations indicating the
existence of another V-type asteroid at $\sim2.53$ AU, (40521) 1999RL95,
and we investigate the possibility that these two asteroids evolved
from the Vesta family to their present orbits by drifting in semi-major
axis due to the Yarkovsky effect. The main problem with this scenario
is that the asteroids need to cross the 3/1 mean motion resonance
with Jupiter, which is highly unstable.

Combining numerical simulations of the orbital evolution, that include
the Yarkovsky effect, with Monte Carlo models, we compute the probability
of an asteroid of given diameter $D$ to evolve from the Vesta family
and to cross over the 3/1 resonance, reaching a stable orbit in the
middle belt. Our results indicate that an asteroid like (21238) 1995WV7
has a low probability (less than 10\%) of having evolved through this
mechanism due to its large size ($D\sim5$ km), because the Yarkovsky
effect is less efficient for larger arteroids. However, the mechanism
might explain the orbit of smaller bodies like (40521) 1999RL95 ($D\sim3$
km), provided that we assume that the Vesta family formed $\gtrsim3.5$
Gy ago. We estimate the debiased population of V-type asteroids that
might exist in the same region as (21238) and (40521) ($2.5<a\lesssim2.62$
AU) and conclude that about 10\% or more of the V-type bodies with
$D>1$ km may come from the Vesta family by crossing over the 3/1
resonance. The remaining 90\% must have a different origin.
\end{abstract}


\newpage{}

Corresponding author:

\vspace{1cm}

Fernando Roig

Rua Gal. José Cristino 77

20921-400, Rio de Janeiro

RJ, BRAZIL

e-mail: froig@on.br

phone: +55 (21) 3878 9205

fax: +55 (21) 2589 8972

\newpage{}

\section{Introduction}

The recent discovery of (21238) 1995WV7 (\citealp{2006DPS....38.7106B};
\citealp{2006Preprint..M}; \citealp{2006astro.ph..9420H}), a basaltic
asteroid in the middle Main Belt ($2.5<a<2.82$ AU), raised new questions
about the origin of basaltic material in the asteroid belt.

Basaltic asteroids show a spectrum characterized by the presence of
a deep absorption band centered at $\sim0.9\ \mu$m, and are classified
as V-type in the usual taxonomies (\citealp{1989aste.conf.1139T};
\citealp{2002Icar..158..146B}). Most of the known V-type asteroids
are fragments from the crust of asteroid (4) Vesta. This is supported
by the fact that (4) Vesta is the only large asteroid showing a basaltic
crust \citep{1970Sci...168.1445M}, and almost all V-type asteroids
are found in the same region of the Main Belt as Vesta, i.e. the inner
belt ($a<2.5$ AU). Moreover, (4) Vesta has associated a dynamical
asteroid family, the Vesta family, whose members are also V-type \citep{2005Icar..174...54M}
and originated from the excavation of a large crater on Vesta's surface
(\citealp{1997Sci...277.1492T}; \citealp{1997M&PS...32..965A}).

The discovery of (1459) Magnya \citep{2000Sci...288.2033L}, a V-type
asteroid in the outer belt ($a>2.82$ AU), provided the first evidence
for another possible source of basaltic asteroids in the Main Belt.
(1459) Magnya is too far away from the Vesta region as to have any
chance of being a fragment from Vesta's crust. No dynamical mechanism
is known to be able to bring an asteroid from the Vesta family to
the Magnya region. Moreover, Magnya is too big (diameter $D\sim17$
km) as to fit within the size distribution of the Vesta family (see
Sect. \ref{sub:sec32}). \citet{2002Icar..158..343M} suggested that
Magnya is the fragment from a differentiated parent body that broke
up in the outer belt, but up to now no other V-type asteroids have
been confirmed in the same region of the belt to support this hypothesis
(\citealp{2007arXiv0704.0230D}; \citealp{2007LPI..38.1663M}).

The case of (21238) 1995WV7, the first V-type asteroid discovered
in the middle belt, has some similarities with the case of Magnya,
but also shows some differences. (21238) 1995WV7 is also far away
from Vesta, and ejection velocities larger than $\sim2$ km/s would
be necessary to directly transport it form Vesta's surface to its
present orbit. These ejection velocities cannot be produced in typical
craterization events similar to the one that originated the Vesta
family \citep{1997M&PS...32..965A}. On the other hand, (21238) 1995WV7
is close to the outer border of the 3/1 mean motion resonance with
Jupiter (hereafter J3/1 MMR), centered at 2.5 AU, and its size ($D\sim5$
km) fits within the size distribution of the Vesta family. Since the
inner border of the J3/1 MMR is very close to the outer edge of the
Vesta family, the possibility of (21238) 1995WV7 being a former member
of this family that reached its present orbit after crossing over
the resonance cannot be totally ruled out. The J3/1 MMR is highly
chaotic \citep{1982AJ.....87..577W} and is considered a difficult-to-cross
barrier, but up to now no detailed studies have been made to confirm
this.

A different scenario is proposed by \citet{2007A&ASS..C}, in which
the source for (21238) 1995WV7 and other, yet undiscovered, basaltic
asteroids in the middle belt could be asteroid (15) Eunomia located
at $a\sim2.65$ AU. This asteroid appears to be a partially differentiated
parent body showing a basaltic-like composition in part of its surface
\citep{2005Icar..175..452N}. Carruba et al. suggest that several
collisions made (15) Eunomia lose its basaltic crust almost completely,
and the subsequent fragments were significantly dispersed in the middle
belt over the age of the Solar System. 

The aim of this work is to study the possible origin of (21238) 1995WV7
and other V-type asteroids in the middle asteroid belt. In particular,
we analyze the possibility that asteroids from the Vesta family increase
their orbital semi-major axis due to the Yarkovsky effect \citep{2000Icar..148..118V}
and cross over the J3/1 MMR reaching stable orbits in the middle belt.
In Section \ref{sec:2}, we introduce the population of V-type asteroids
observed by the Sloan Digital Sky Survey in the middle belt, and present
our spectroscopic observations with the Gemini South Telescope that
allow to confirm another V-type asteroid in the same region as (21238).
In Section \ref{sec:3}, we describe our simulations, analyze the
resonance crossing mechanism proposed above, and evaluate its efficiency
to produce V-type asteroids beyond 2.5 AU. In Section \ref{sec:4},
we discuss our results compared to the debiased distribution of V-type
asteroids in the middle belt. Finally, Section \ref{sec:5} is devoted
to the conclusions.

\section{V-type asteroids in the middle belt\label{sec:2}}

The existence of V-type asteroids in the middle belt was first predicted
by \citet{2006Icar..183..411R}, who analyzed the colors of the Sloan
Digital Sky Survey Moving Objects Catalog (SDSS-MOC; \citealp{2001AJ....122.2749I}).
These authors studied a sample of 13\,290 asteroids contained in
the 3rd release of the SDSS-MOC that show errors smaller than 10\%
in all the five bands of the SDSS photometric system, named $u,g,r,i,z$,
respectively. They found three candidate V-type asteroids in the middle
belt, which are listed in Table \ref{tab:1}. Two of them are located
very close to the outer border of the J3/1 MMR. The third one is close
to the outer border of the J8/3 MMR, centered at 2.7 AU. The spectroscopic
confirmation of the basaltic nature of (21238) was reported by \citet{2006DPS....38.7106B}
and \citet{2006Preprint..M}, based on spectroscopic observations
in the near infrared (NIR), and also by \citet{2006astro.ph..9420H}
based on visible spectrocopic plus NIR photometry.

\begin{table}[th]
\caption{Proper semi-major axis $a_{p}$, proper eccentricity $e_{p}$, sin
of proper inclination $\sin I_{p}$, absolute magnitude $H$ and diameter
$D$ of predicted V-type asteroids in the middle belt according to
the colors of the SDSS-MOC. Diameters have been estimated assuming
and albedo of 0.4, similar to the albedo of (4) Vesta (0.42, \citealp{1989aste.conf.1090T}).}
\label{tab:1}\bigskip{}

\noindent \begin{centering}
\begin{tabular}{lccccc}
\hline 
Name  & $a_{p}$ {[}AU]  & $e_{p}$  & $\sin I_{p}$  & $H$  & $D$ {[}km]\tabularnewline
\hline 
(21238) 1995WV7  & 2.54108  & 0.1371  & 0.1866  & 13.04  & 5.15\tabularnewline
(40521) 1999RL95  & 2.53111  & 0.0458  & 0.2159  & 14.36  & 2.80\tabularnewline
(66905) 1999VC160  & 2.74627  & 0.1457  & 0.2291  & 14.51  & 2.62\tabularnewline
\hline
\end{tabular}
\par\end{centering}
\end{table}

As part of an observational campaign to confirm the taxonomy of V-type
candidates identified by \citet{2006Icar..183..411R}, we obtained
spectra of (21238) and (40521) in the visible. The observations were
carried out during the nights of 29-30 April, 2006, at the Gemini
South Observatory (GS), using the Gemini Multi-Object Spectrograph
(GMOS). Table \ref{tab:2} provides the observational circumstances
of the targets. In order to remove the solar signature from the asteroids
spectra, we also observed the stars SA 107-871 ($V=12.4$) and SA
110-361 ($V=12.4$), taken form the Selected Areas of \citet{1992AJ....104..340L},
that we used as solar analog stars%
\footnote{These are G2V stars with solar colors in the magnitude range allowed
by Gemini telescopes ($V>11$). We recall that no spectroscopic solar
analog stars are available in the literature in this magnitude range.%
}.

\begin{table}[th]
\caption{Observational circumstances of (21238) and (40521): $\Delta$ is geocentric
distance in AU, $r$ is heliocentric distance in AU, $\phi$ is solar
phase angle, $\theta$ is solar elongation, $V$ is visible magnitude.
Dates correspond to the starting time of the observations. The observing
conditions were better than 85\% of image quality, 70\% of sky transparency
(cloud cover), 100\% of sky transparency (water vapor), 80\% of sky
background, air mass $<1.5$.}
\label{tab:2}\bigskip{}

\noindent \begin{centering}
\begin{tabular}{ccccccccc}
\hline 
{\small Asteroid}  & {\small Date {[}UT]}  & {\small $\alpha$ (J2000)}  & {\small $\delta$ (J2000)}  & {\small $\Delta$}  & {\small $r$}  & {\small $\phi$}  & {\small $\theta$}  & {\small $V$}\tabularnewline
\hline 
{\small (21238)}  & {\small 2006 Apr 29.4094}  & {\small $20^{\mathrm{h}}44^{\mathrm{m}}53.01^{\mathrm{s}}$}  & {\small $-24^{\circ}57^{\prime}12.8^{\prime\prime}$}  & {\small 2.599}  & {\small 2.818}  & {\small $20.9^{\circ}$}  & {\small $91.9^{\circ}$}  & {\small 18.3}\tabularnewline
{\small (40521)}  & {\small 2006 Apr 30.2811}  & {\small $14^{\mathrm{h}}11^{\mathrm{m}}29.86^{\mathrm{s}}$}  & {\small $-24^{\circ}01^{\prime}44.8^{\prime\prime}$}  & {\small 1.489}  & {\small 2.487}  & {\small $4.1^{\circ}$}  & {\small $169.8^{\circ}$}  & {\small 18.1}\tabularnewline
\hline
\end{tabular}
\par\end{centering}
\end{table}

Tracking of the asteroids at non-sideral rate was not possible because
the use of the peripherical Wavefront Sensor (WFS) is not recommended
due to flexure within GMOS. Instead, we used the On-Instrument Wavefront
Sensor (OIWFS) tracking at sideral rate, with the slit oriented in
the direction of the asteroid's proper motion.

All the observations were performed using the following GMOS configuration:
grating R400, filter OG515 to avoid second order spectrum contamination
longwards of $\sim0.7\,\mu$m, slit width 1.5 arcsec (the maximum
allowable with GS-GMOS), central wavelength $0.73\,\mu$m, spectral
coverage $0.522$ to $0.938\,\mu$m, spatial binning 2, and spectral
binning 4. This configuration provides a resolution $R\sim3\,000$
at $0.90\,\mu$m, but $R\sim200$ is enough to detect the deep absorption
band longwards of $0,75\,\mu$m typical of V-type spectra. Therefore,
it was possible to do a 15:1 rebinning of the asteroids spectra to
improve the final signal-to-noise (S/N) ratio by a factor of $\sim4$.

Each asteroid was observed six times at six different positions along
the slit separated by 10 arcsec. Each solar analog star was observed
three times at three different positions along the slit with the same
separation. The integration times for the asteroids allowed to attain
S/N$\sim20$ at $0.90\,\mu$m which, after spectral rebinning, provided
a minimum S/N $\sim70$. Table \ref{tab:3} summarizes the asteroids
exposure times. The exposure times for the solar analog stars were
chosen to reach 50-70\% of the detector full well (100k electrons)
per exposure. Wavelength calibration of the spectra was performed
using a standard CuAr lamp.

\begin{table}[th]
\caption{Number of exposures, exposure times (individual and total), S/N at
$0.90\,\mu$m of the raw and rebinned spectra, and solar analog used
in the reduction.}
\label{tab:3}\bigskip{}

\noindent \begin{centering}
\begin{tabular}{ccccccc}
\hline 
Asteroid  & $n_{\mathrm{exp}}$  & $T_{\mathrm{exp}}$  & $T_{\mathrm{total}}$  & S/N (raw)  & S/N (rebinned)  & Solar analog\tabularnewline
\hline 
(21238)  & 6  & 200 sec  & 1200 sec  & 20  & 80  & SA 110-361\tabularnewline
(40521)  & 6  & 500 sec  & 3000 sec  & 18  & 72  & SA 107-871\tabularnewline
\hline
\end{tabular}
\par\end{centering}
\end{table}

The GMOS IRAF package was used to perform the standard reduction tasks.
Since each image was dithered along the spatial direction, we remove
the fringes by making a background image resulting from the combination
of three successive images centered around the time of the image we
want to correct. After the fringe correction, the individual frames
were coadded to improve the S/N and the spectrum was extracted.

\begin{figure}[th]
\caption{Visible spectra of (21238) and (40521) observed with GMOS at Gemini
South (black lines). The gray lines are the spectra of other known
V-type asteroids. The spectra are normalized to 1 at 5500 $\textrm{\AA}$
and shifted by 0.5 in reflectance for clarity. A running box of 20
$\textrm{\AA}$ has been applied to smooth the spectra.}
\label{fig:1}\bigskip{}

\noindent \begin{centering}
\includegraphics[clip,width=1\columnwidth]{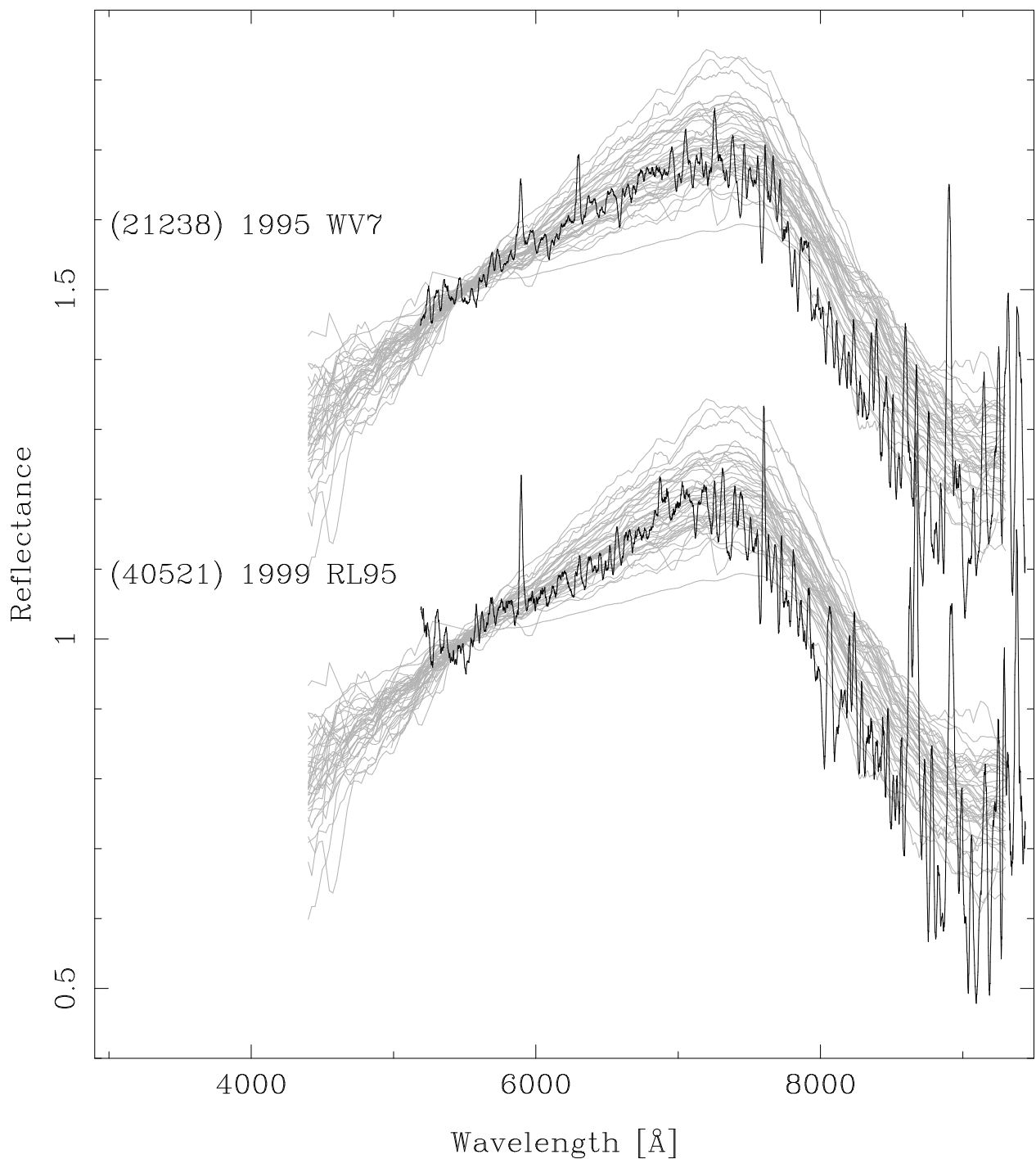}
\par\end{centering}
\end{figure}

The final spectra of (21238) and (40521) are shown in Figure \ref{fig:1}.
We have plot in gray several spectra of known V-type asteroids taken
from the SMASS \citep{2002Icar..158..106B} and S$^{3}$OS$^{2}$
\citep{2004Icar..172..179L} surveys for comparison. Both spectra
are compatible with the V class, showing the typical absorption band
with a minimum at $0.90\,\mu$m. Our observations with Gemini constitute
another independent confirmation of the basaltic nature of (21238),
and also allow the confirmation of (40521) as the second V-type asteroid
found in the middle belt%
\footnote{The ASCII files of the spectra are available at \url{http://staff.on.br/froig/vtypes}%
}.

\section{Possible origin of V-type asteroids in the middle belt\label{sec:3}}

In the following, we analyze the possibility that (21238) and (40521)
were fragments of (4) Vesta that evolved to their current orbits by
the Yarkovsky effect. The main problem with this scenario is that
these asteroids needed to cross the J3/1 MMR which is notably unstable.
To estimate the time $\Delta t_{\mathrm{cross}}$ that a 5 km asteroid
would require to cross the J3/1 MMR, \citet{2006Preprint..M} and
\citet{2006astro.ph..9420H} divided the resonance width $\Delta a$
by the drift rate $da/dt$ due to the Yarkovsky effect, and concluded
that $\Delta t_{\mathrm{cross}}\gg t_{\mathrm{inst}}$, where $t_{\mathrm{inst}}$
is the instability time scale of the J3/1 MMR (i.e., the time required
for a population of asteroids to be removed from the resonance, see
\citealp{1997Sci...277..197G}). These authors thus inferred that
it would be impossible for an asteroid like (21238) to cross the J3/1
MMR. However, this argument is only approximative because it does
not take into account the resonance dynamics. Here we used precise
numerical simulations to study whether the crossing is actually possible.

We performed a series of simulations of the evolution of test particles
representing selected members of the Vesta dynamical family. To define
this family, we used the database of $188\,207$ asteroid proper elements
released by March 2005%
\footnote{AstDys \url{http://hamilton.dm.unipi.it/cgi-bin/astdys/astibo}. This
database is contemporary to the 3rd. release of the SDSS-MOC, therefore
the two datasets can be directly compared.%
}. We applied the Hierarchical Clustering Method (HCM - \citealp{1990AJ....100.2030Z})
and defined the Vesta family at a cutoff of 60 m/s, which is 5 m/s
larger than the predicted quasi-random level%
\footnote{The average minimum distance between pairs of neighbor asteroids.%
} for the inner belt. This guaranteed that we defined the Vesta family
with the largest possible number of members ($\sim9\,500$) that could
be detected from the given dataset of proper elements.

Among the members of the Vesta family, we selected 21 asteroids with
$2.485\leq a_{p}\leq2.490$ AU, which are the closest ones to the
J3/1 MMR, and generated 100 clones of each. All 100 clones had the
same orbital parameters as the original asteroid, but we allowed each
to drift in semi-major axis at a slightly different rate $da/dt>0$
due to the Yarkovsky effect. This implied that the clones reached
the border of the J3/1 MMR at slightly different phases of the resonant
angle $\sigma=3\lambda_{\mathrm{J}}-\lambda-\varpi$, where $\lambda_{\mathrm{J}},\lambda$
are the mean longitudes of Jupiter and the asteroid, respectively,
and $\varpi$ is the longitude of perihelion of the asteroid. Therefore,
they sampled different resonant interaction histories.

The simulations were performed using a modified version of the SWIFT\_MVS
integrator \citep{1994Icar..108...18L}. The bodies were considered
as massless particles subject to perturbations from all planets except
Mercury, and the Yarkovsky effect was introduced in the simulation
as an additional acceleration term depending on the velocity as:\[
\left(\frac{d^{2}\vec{r}}{dt^{2}}\right)_{\mathrm{Yarko}}=\frac{GM}{2a^{2}}\ \frac{da}{dt}\ \frac{\vec{v}}{v^{2}}\]
where $G$ is the gravitational constant, $M$ the mass of the Sun,
$a$ the osculating semi-major axis of the orbit, $\vec{r},\vec{v}$
the instantaneous position and velocity of the body, and $da/dt$
the required drift rate measured in AU/y. According to the analytical
theory of \citet{1999A&A...344..362V}, the Yarkovsky drift rate approximately
scales with diameter as:\begin{equation}
\frac{da}{dt}\simeq2.5\times10^{-4}\ \frac{1\ \mathrm{km}}{D}\ \cos\epsilon\label{eq:1}\end{equation}
where $D$ is in km, $\epsilon$ is the spin axis obliquity, and the
coefficient was obtained assuming physical and thermal parameters
typical of basalt and albedo 0.4. We chose the thermal parameters
to produce large but plausible drift rates, because this would favor
the J3/1 MMR crossing. Slower drift rates would apply if the real
thermal parameters have different values. The so-called seasonal Yarkovsky
effect \citep{1995JGR...100.1585R} was not included in our model
because it only produces $da/dt<0$ and it is an order of magnitude
smaller than the diurnal effect modeled by Eq. (\ref{eq:1}) for km-size
asteroids. The collisional reorientation of spin axes \citep{1979Icar...40..145H}
and the YORP effect \citep{2000Icar..148....2R} were not taken into
account either. The former effect has proven not to be so relevant
in changing the spin axis obliquities as believed \citep{2006AREPS..34..157B},
while the YORP effect statistically tends to align the spin axes around
$\epsilon\sim0,\pi$, thus forcing a maximum Yarkovsky drift, on average,
for all the particles. 

We perfomed two series of simulations. In the first series, we assumed
that each clone drifted outwards at a rate randomly chosen in the
interval $(1.0\pm0.1)\times10^{-4}$ AU/My, that according to Eq.
(\ref{eq:1}) corresponds to $D\sim2.5$ km V-type asteroids. These
simulations spanned 150 My. In the second series, the clones drifted
at a rate randomly chosen in the interval $(1.0\pm0.1)\times10^{-3}$
AU/My, which corresponds to $D\sim250$ m asteroids. These simulations
spanned 25 My. The time step of the integrator was set to be 0.04
y.

In both simulations, more than $\sim97$\% of the particles were discarded
because they entered the J3/1 MMR and chaotically evolved to planet
crossing orbits in a few My. These particles were eliminated by close
encounters with the planets, mainly the Earth and Mars, or by impacting
the Sun. But a small fraction --less than $\sim3$\%-- of the particles
crossed over the J3/1 MMR and ended the simulation in stable orbits
with $a>2.5$ AU.

\begin{figure}[th]
\caption{Evolution of the semi-major axis, eccentricity and inclination of
two test particles crossing over the J3/1 MMR at two different Yarkovsky
drift rates. Note the different scales in the time axis.}
\label{fig:2}\bigskip{}

\begin{centering}
\includegraphics[clip,width=1\textwidth]{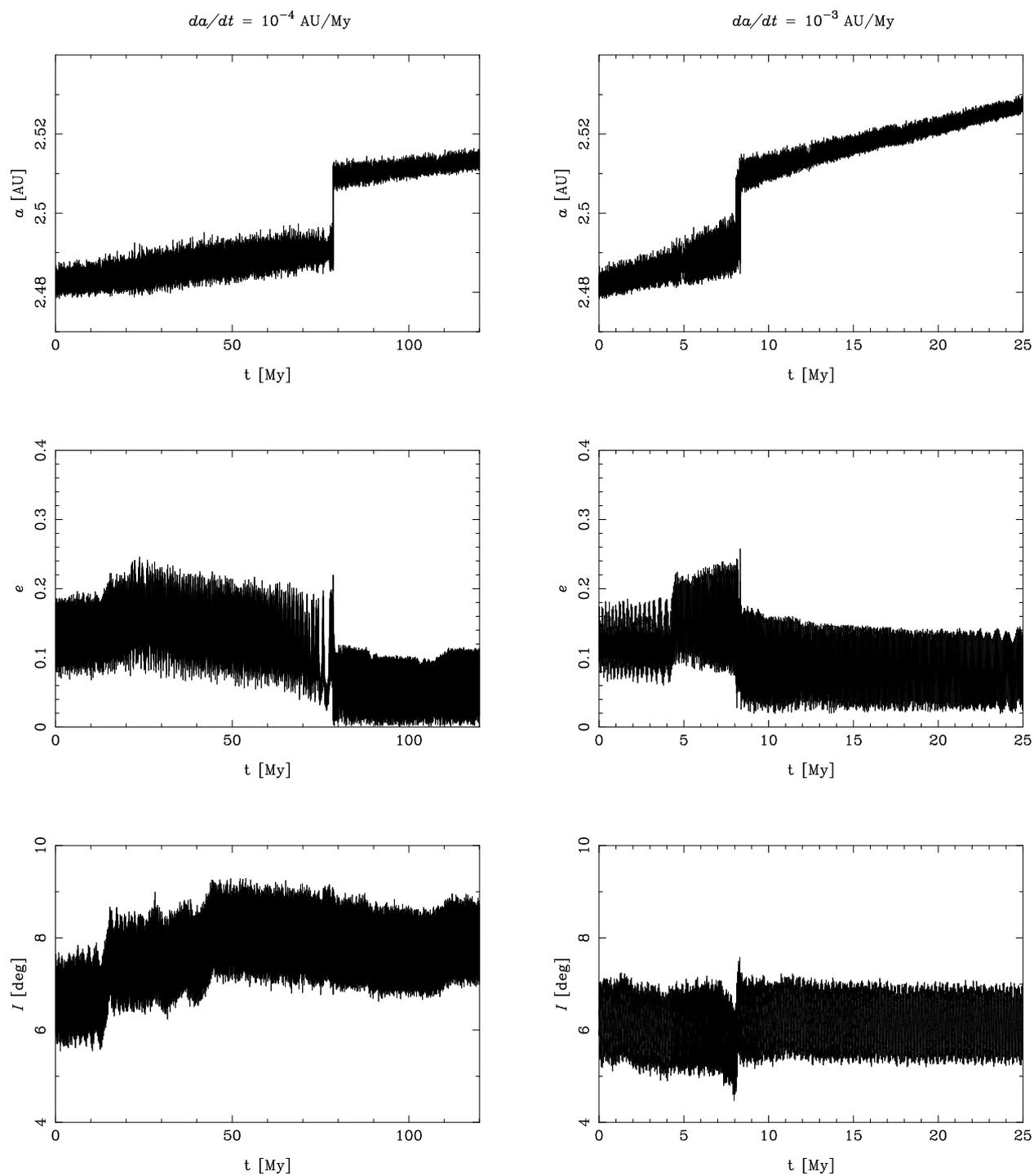}
\par\end{centering}
\end{figure}

Figure \ref{fig:2} shows two examples of test particles that crossed
over the J3/1 MMR. The left panels correspond to a slowly drifting
particle and the right panels correspond to a faster drift. The particles
enter the J3/1 MMR at the side of lower semi-major axes, perform a
few librations around the resonant semi-major axis, and exit the resonance
at the side of larger semi-major axes. The particles remain in the
resonance for at most a few $10^{4}$ y. Their eccentricities and
inclinations are not significantly affected by the passage through
the resonance. In the examples shown in Fig. \ref{fig:2}, the eccentricities
decay to lower values but this happens \textit{after} the particles
have already crossed the J3/1 MMR%
\footnote{This effect is probably due to the interplay with non linear secular
resonances located close to the border of the J3/1 MMR, like the $g-2g_{6}+g_{5}$
resonance, where $g$ is the secular frequency of the asteroid perihelion,
and $g_{5},g_{6}$ are the frequencies of the perihelia of Jupiter
and Saturn, respectively.%
}. In general, the particles that jumped the J3/1 MMR ended the simulations
either with higher or lower values of the eccentricities and inclinations
than their original values.

\begin{figure}[th]
\caption{Detail of the trajectory during the resonance crossing of the slowly
drifting test particle shown in Fig. \ref{fig:2}. The top panel shows
the resonant angle $\sigma$ as a function of time. The bottom panel
show the trajectory in the space $a,\sigma$. The arrows indicate
the direction of the trajectory. The numbers provide the temporal
sequence of the trajectory.}
\label{fig:3}\bigskip{}

\noindent \begin{centering}
\includegraphics[bb=35bp 300bp 415bp 840bp,clip,height=0.9\textheight]{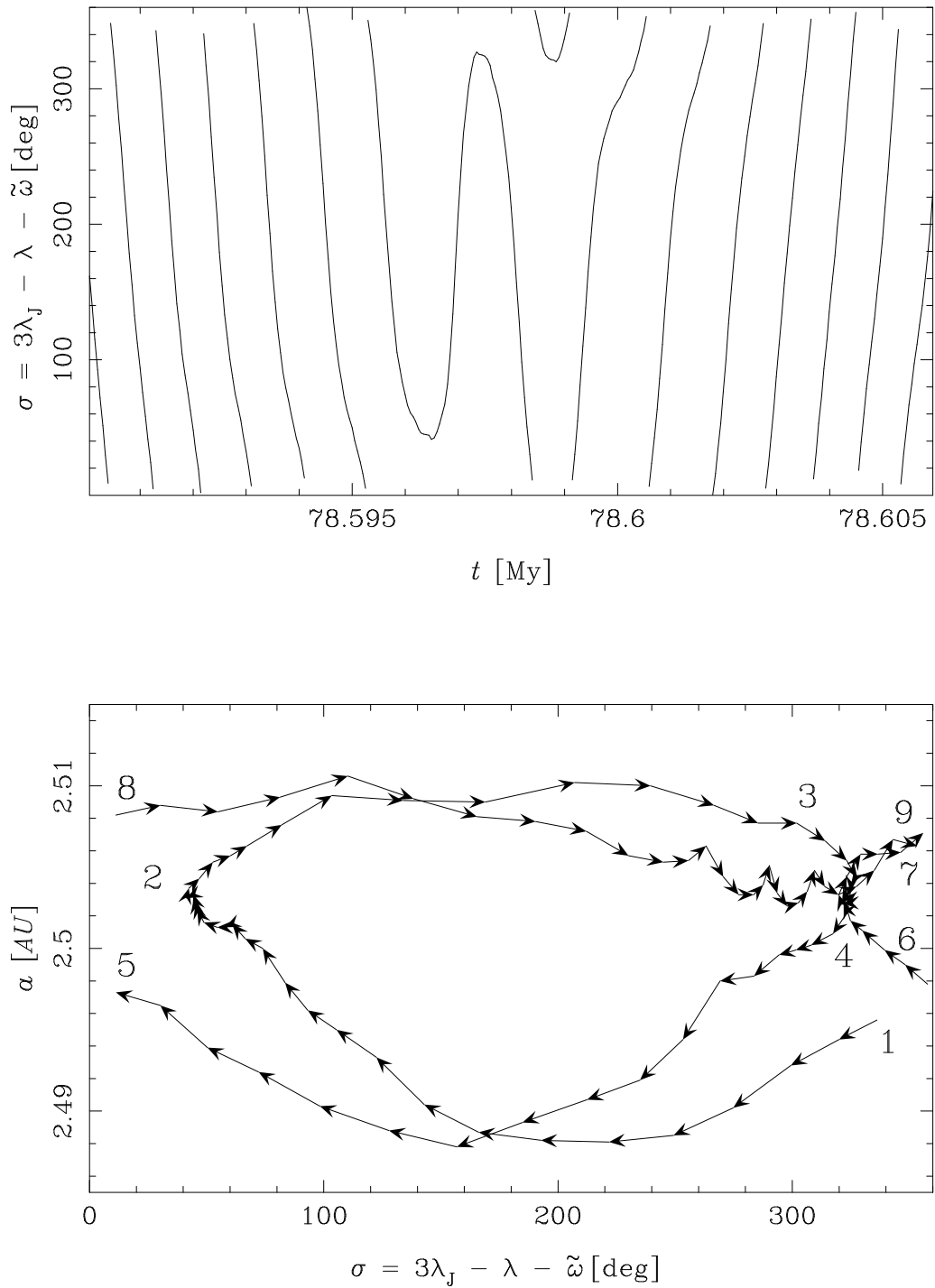}
\par\end{centering}
\end{figure}

In Fig. \ref{fig:3} we show the trajectory of the slowly drifting
particle of Fig. \ref{fig:2} at the exact moment in which it crosses
the J3/1 MMR. We can see that the particle spent only $\sim5\,000$
y inside the resonance (top panel), so it virtually ``jumps'' over
the resonance. The numbers in the bottom panel provide the temporal
sequence of the trajectory. The particle enters the resonance somewhere
between ``1'' and {}``2'', then it performs one and a half libration
sticking close to the resonance separatrix, and finally exits the
resonance somewhere between ``8'' and {}``9''. This trajectory is
a typical example of the resonance crossing mechanism. We expect that
only the orbits that remain close to the separatrix would be able
to exit the resonance pushed by the Yarkovsky drift, and this might
depend on the particular phase of the resonant angle, $\sigma$, that
the orbit has when it enters the resonance.

\begin{table}[th]
\caption{Fraction $f_{\mathrm{cross}}$ of asteroids that crossed the J3/1
MMR during our simulations for different Yarkovsky drifts.}
\label{tab:6}\bigskip{}

\noindent \begin{centering}
\begin{tabular}{cc}
\hline 
$da/dt$ $\times10^{-4}$ {[}AU/My]  & $f_{\mathrm{cross}}$\tabularnewline
\hline 
$1.0\pm0.1$  & $0.0033\pm0.0019$\tabularnewline
$10.0\pm1.0$  & $0.029\pm0.005$\tabularnewline
\hline
\end{tabular}
\par\end{centering}
\end{table}

From our simulations, it was possible to estimate the fraction of
test particles that crossed the J3/1 MMR, $f_{\mathrm{cross}}$, and
its formal uncertainty, as a function of the Yarkovsky drift rate.
The results are given in Table \ref{tab:6}. Assuming that these results
reflect a linear dependence between $f_{\mathrm{cross}}$ and $da/dt$
within the studied range of drift rates, and taking into account Eq.
(\ref{eq:1}), we find that\begin{equation}
f_{\mathrm{cross}}(D)\simeq0.0075\ \frac{1\ \mathrm{km}}{D}\ \cos\epsilon.\label{eq:2}\end{equation}
The application of this formula in the size range $1\mathrm{\, km}\lesssim D\lesssim5\mathrm{\, km}$
requires only modest extrapolations of the linear dependence toward
smaller values of the drift rate ($\sim0.5\times10^{-4}$ AU/My).
This result demonstrates, for the first time, that the crossing of
Vesta family members over the J3/1 MMR is possible in practice. The
crossing probability ranges from 0.15\% to 0.25\% for asteroids in
the size range of (21238) and (40521).

\section{The production of V-type asteroids with $a>2.5$ AU\label{sub:sec32}}

In the preceding section, we have shown that it is possible for a
former Vesta family member to cross over the J3/1 MMR, reaching a
stable orbit in the middle belt. We have also estimated the crossing
probability as a function of the size (Eq. \ref{eq:2}) to be $\sim0.1$-0.8\%
for km-size asteroids. We may now wonder whether this may actually
explain the presence of (21238) and (40521) in the middle belt.

The number of Vesta family members, with diameters between $D$ and
$D+dD$, that cross over the J3/1 MMR during the age of the Vesta
family, $\tau_{\mathrm{age}}$, and end in the middle belt is: \begin{equation}
dN_{a>2.5}(D)=dN_{0}(D)\, f_{\mathrm{col}}(D)\, f_{\mathrm{reach}}(D)\, f_{\mathrm{cross}}(D)\label{eq:3}\end{equation}
where $dN_{0}(D)$ is the number of Vesta family members that were
produced when the family formed. The factor $f_{\mathrm{col}}(D)$
accounts for the fraction of Vesta family members in the interval
$[D,D+dD]$ that, on one hand, survived and, on the other hand, were
created due to collisional evolution in $\tau_{\mathrm{age}}$. It
is worth noting that the product $dN_{0}(D)\, f_{\mathrm{col}}(D)$
is equivalent to the presently observed number of family members,
$dN(D)$. The factor $f_{\mathrm{reach}}(D)$ accounts for the fraction
of Vesta family members that were able to reach the border of the
J3/1 MMR in $\tau_{\mathrm{age}}$ aided by the Yarkovsky effect.
Finally, the factor $f_{\mathrm{cross}}(D)$ is provided by Eq. (\ref{eq:2}).
In order to apply Eq. (\ref{eq:3}), we have to address three main
issues.

The first issue concerns the actual age of the Vesta family, $\tau_{\mathrm{age}}$.
Initial estimates by \citet{1996A&A...316..248M}, based on collisional
evolution models aiming to reproduce the presently observed size frequency
distribution of the Vesta family, indicated $\tau_{\mathrm{age}}\sim1$-2
Gy. However, these estimates have large uncertainties due to unknown
initial conditions and to uncertainties in the various collisional
parameters. \citet{2005A&A...441..819C}, based on the dynamical evolution
of some Vesta family {}``fugitives'', found that $\tau_{\mathrm{age}}\gtrsim1.2$
Gy. On the other hand, \citet{2003M&PS...38..669B}, measuring isotopic
abundances in the Howardite-Eucrite-Diogenite (HED) meteorites --that
presumably come from Vesta, \citep{1997M&PS...32..903M}--, concluded
that Vesta's crust suffered several major craterization impacts $\gtrsim3.5$
Gy ago. This could imply $\tau_{\mathrm{age}}\gtrsim3.5$ Gy. In view
of the wide range of possible ages, we will adopt for our calculations
three different values of $\tau_{\mathrm{age}}$: 1.5, 2.5 and 3.5
Gy.

The second issue to address is to estimate the fraction of family
members, $f_{\mathrm{reach}}$, that would be able to reach the J3/1
MMR in $\tau_{\mathrm{age}}$ due to the Yarkovsky effect. For this
purpose, we used a Monte Carlo algorithm. We generated an artificial
Vesta family constituted of $10\,000$ fragments with individual ejection
velocities, $v_{\mathrm{ej}}$, attributed by assuming a Maxwellian
distribution with mean ejection velocity $\bar{v}_{\mathrm{ej}}$%
\footnote{We recall that the specific energy of the impact that generated the
family is $\propto\bar{v}_{\mathrm{ej}}^{2}$.%
}. We considered only the fragments with \[
v_{\mathrm{ej}}^{2}-v_{\mathrm{esc}}^{2}>0\qquad\qquad\mathrm{and}\qquad\qquad v_{\mathrm{ej}}^{2}-v_{\mathrm{cut}}^{2}<0\]
where $v_{\mathrm{esc}}=314$ m/s is the escape velocity from the
surface of (4) Vesta%
\footnote{$v_{\mathrm{esc}}=\sqrt{1.64\ G\,\frac{4}{3}\pi\rho R^{2}}$ being
$R$ the radius of Vesta and $\rho$ its density \citep{1993cemda..57....1p}.%
}, and $v_{\mathrm{cut}}=600$ m/s is a maximum cutoff velocity \citep{1997M&PS...32..965A}.
We further assumed that, at the moment of the impact, (4) Vesta had
a true anomaly and perihelion argument such that the final distribution
of the fragments in proper elements space is spread over a wide range
of $a_{p},e_{p},I_{p}$ \citep{1995Icar..118..132M}. Finally, we
considered that all the fragments had the same diameter --i.e., the
ejection velocity does not depend on size--, and attributed to each
fragment a random value of $\cos\epsilon$ between $-1$ and 1. Using
Eq. (\ref{eq:1}), we computed for each fragment the total drift over
$\tau_{\mathrm{age}}$ and determined the fraction $f_{\mathrm{reach}}$
that ended with $a>2.5$ AU.

\begin{figure}[th]
\caption{The fraction $f_{\mathrm{reach}}$ of Vesta's fragments that reached
the J3/1 MMR border along three different values of $\tau_{\mathrm{age}}$,
and for two different values of $\bar{v}_{\mathrm{ej}}$: 150 m/s
(full lines) and 300 m/s (dashed lines).}
\label{fig:4}\bigskip{}

\noindent \begin{centering}
\includegraphics[clip,width=1\columnwidth]{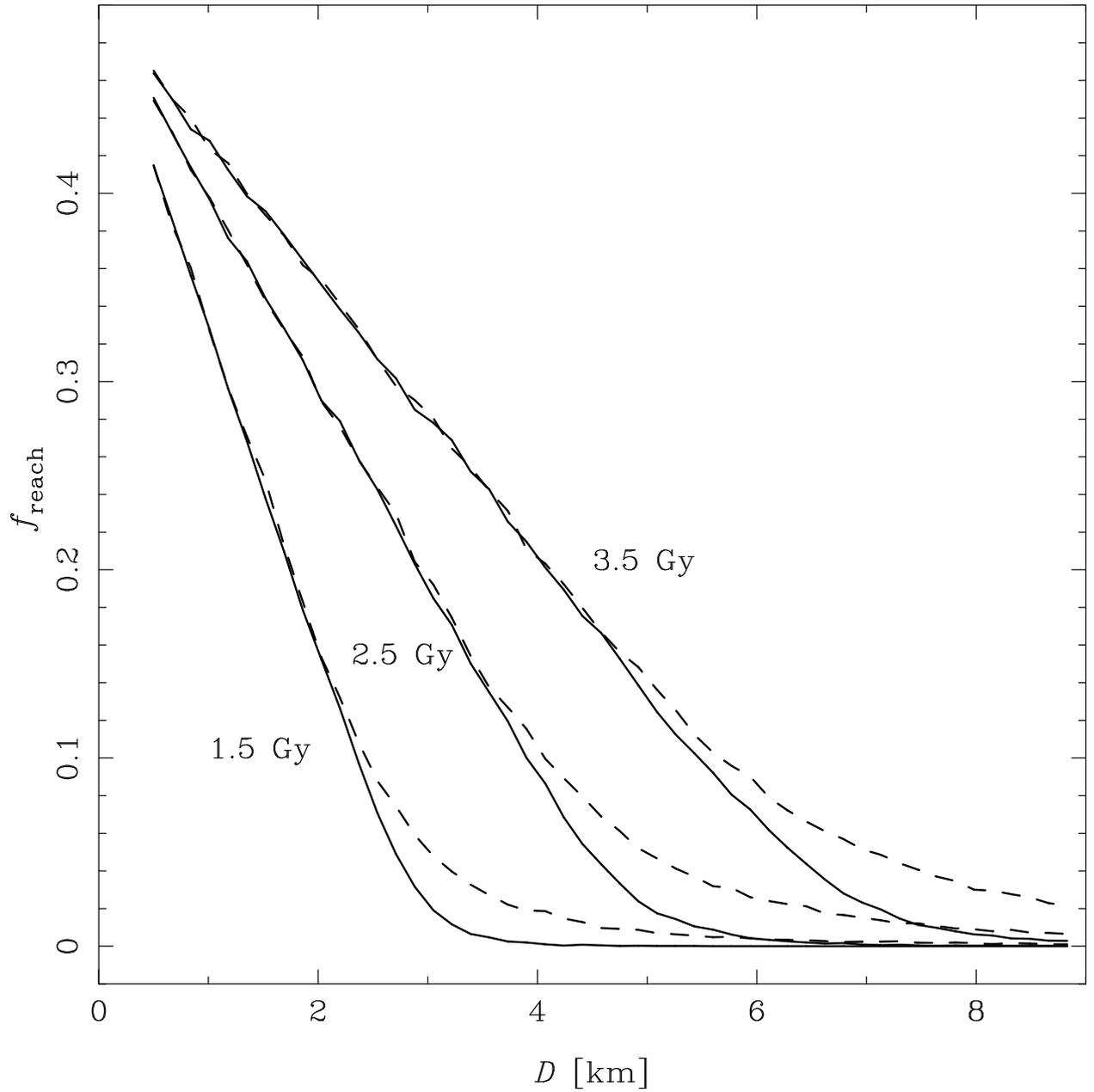}
\par\end{centering}
\end{figure}

Figure \ref{fig:4} shows $f_{\mathrm{reach}}(D)$ as a function of
$\tau_{\mathrm{age}}$ and $\bar{v}_{\mathrm{ej}}$. For very small
sizes, $f_{\mathrm{reach}}$ weakly depends on $\bar{v}_{\mathrm{ej}}$
and tends to 50\% for $D\rightarrow0$. This means that about half
of the smallest fragments will eventually reach the J3/1 MMR (the
other half has $\cos\epsilon<0$ so is drifting inwards). On the other
hand, for very large sizes $f_{\mathrm{reach}}\rightarrow0$ since
large bodies are much less affected by the Yarkovsky effect. It is
worth noting that size range where the behavior of $f_{\mathrm{reach}}$
is more critical, i.e. more sensitive to the different parameters
and especially to the family age, is between 2.0 and 7.0 km, which
is precisely the size range of the asteroids listed in Table \ref{tab:1}. 

The third issue we need to address concerns the determination of the
size frequency distribution (SFD), $dN(D)$, of the Vesta family.
The Vesta family, as defined by the HCM (Section \ref{sec:3}), shows
a very peculiar SFD. Going from larger to smaller sizes, we first
find (4) Vesta with $D\sim460$ km. Then we find four asteroids with
$25\lesssim D\lesssim130$ km and twelve asteroids with $8\lesssim D\lesssim15$
km, which taxonomic types, based on spectroscopic observations (\citealp{2002Icar..158..146B};
\citealp{2004Icar..172..179L}), are different from the V-type. These
asteroids are interlopers in the family and we can exclude them. Finally,
we find a huge amount of members with $D<8$ km. This group includes
all the vestoids, i.e. the known V-type asteroids that are members
of the Vesta family. According to simulations of asteroid fragmentation
using hydrocodes \citep{2007Icar..186..498D}, this kind of SFD, where
there is only one large asteroid and a huge amount of very small fragments
with no bodies in between, would be typical of craterization events
as the one that presumably formed the Vesta family.

\begin{figure}[th]
\caption{The cumulative size distribution of the Vesta family after subtracting
the known interlopers (full line) and three power law fits for different
size ranges (dashed lines). Diameters were computed assuming albedo
0.4 for all bodies.}
\label{fig:5}\bigskip{}

\noindent \begin{centering}
\includegraphics[clip,width=1\columnwidth]{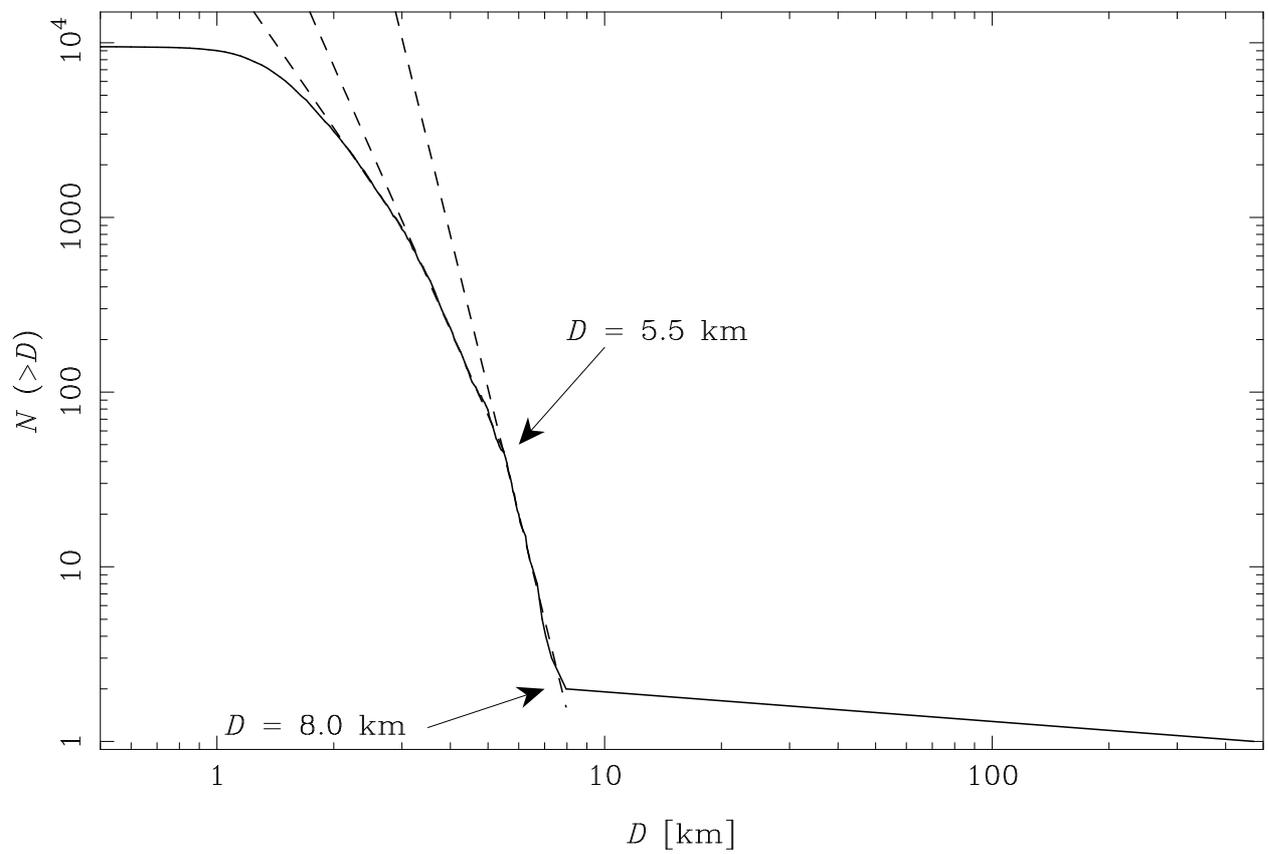}
\par\end{centering}
\end{figure}

Now, $dN(D)$ is related to the differential SFD, $n(D)$, through\[
dN(D)=n(D)\, dD\]
and the cumulative SFD, i.e., the number of family members with diameter
$>D$, is given by \begin{equation}
N(D)=\int_{D}^{D_{\mathrm{max}}}n(D)\, dD.\label{eq:4}\end{equation}
where $D_{\mathrm{max}}$ is the size of the largest family member
(i.e. (4) Vesta). Figure \ref{fig:5} shows the cumulative SFD of
the Vesta family after removing the known interlopers. For $D<8$
km the cumulative SFD can be fit to power laws of the form $N_{0}D^{\gamma}$
(dashed lines in Fig. \ref{fig:5}), where $\gamma$ and $N_{0}$
adopt different values at different size ranges. This change in $\gamma$
and $N_{0}$ is produced by two different effects:

\begin{enumerate}
\item The fact that the sample of known family members is complete only
up to a given size. It is usually assumed that this completeness limit
corresponds to the size where the cumulative SFD shows the first inflexion
point (e.g. \citealp{1999Icar..141...65T}), which is 5.5 km in Fig.
\ref{fig:5}.
\item The natural dynamical/collisional evolution of the family members,
which is known to produce a shallow cumulative SFD at small sizes
\citep{2003Icar..162..328M}. 
\end{enumerate}
Since both these effects are difficult to estimate, we will adopt
here two extreme approaches to model the cumulative SFD:

\begin{itemize}
\item To use the observed cumulative SFD as shown in Fig. \ref{fig:5}.
This model partly accounts for effect \#2, but has the drawback that
it will be biased by the incompleteness of the sample for $D<5.5$
km.
\item To use a single power law fit between 8 and 5.5 km and to extrapolate
it to smaller sizes. This model partly accounts for effect \#1, but
has the drawback that it will significantly overestimate the number
of small members in the Vesta family, especially in the size range
2.0-5.5 km that is crucial for our study. With this approach, the
differential SFD will also be modeled by a single power law of the
form $n_{0}D^{\gamma-1}$.
\end{itemize}
\begin{figure}[th]
\caption{The cumulative distribution $N_{a>2.5}$ for three different values
of $\tau_{\mathrm{age}}$ and two models of the SFD: the observed
SFD (full lines) and the single power law SFD (dashed lines). In all
cases $\bar{v}_{\mathrm{ej}}=200$ m/s. The stars represent the known
V-type asteroids in the middle belt.}
\label{fig:6}\bigskip{}

\noindent \begin{centering}
\includegraphics[clip,width=1\columnwidth]{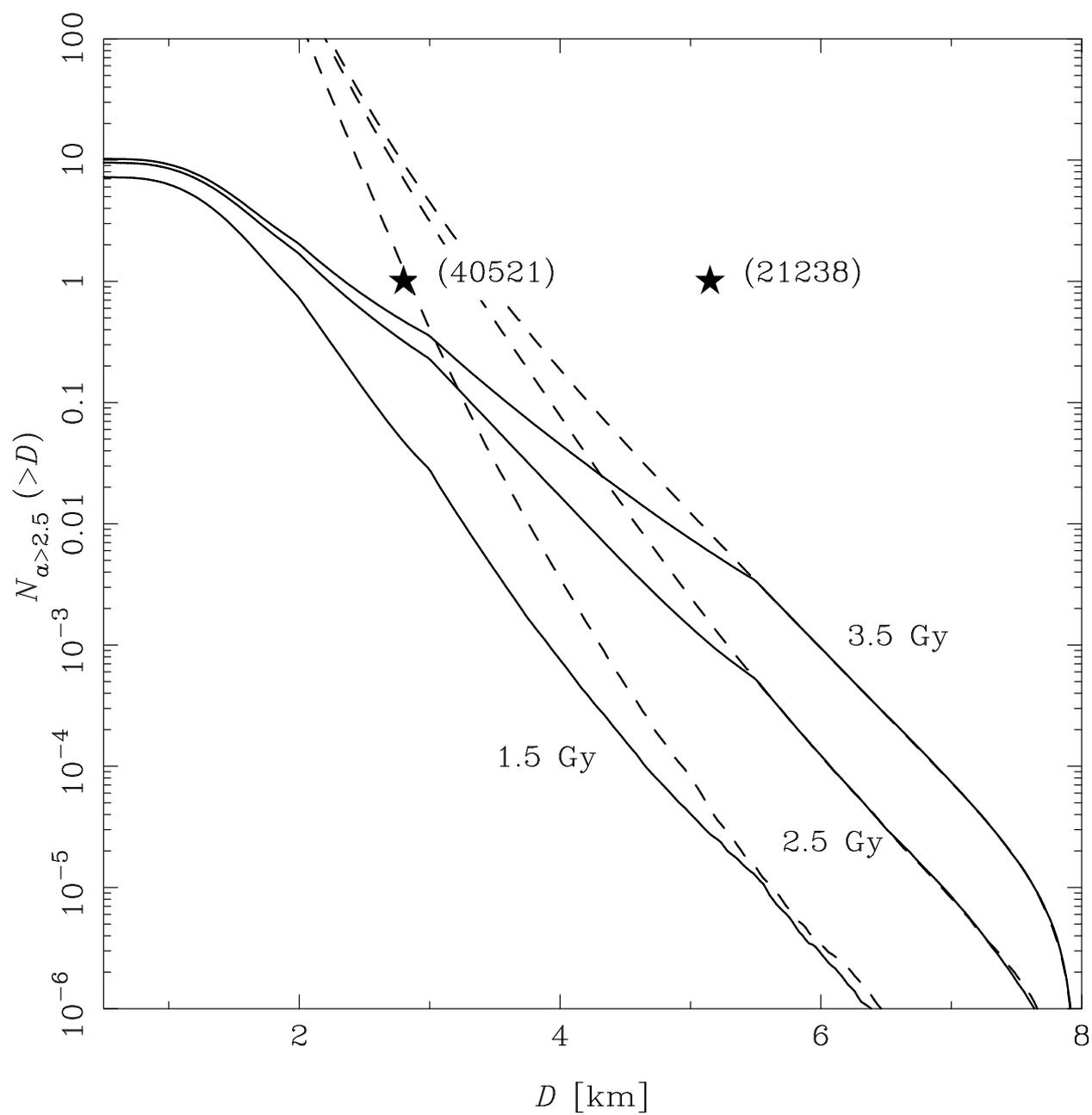}
\par\end{centering}
\end{figure}

Knowing all the quantities involved in Eq. (\ref{eq:3}), it is now
possible to determine the number of Vesta family members with size
$>D$ that cross over the J3/1 MMR and reach the middle belt in $\tau_{\mathrm{age}}$:

\begin{equation}
N_{a>2.5}(D)=\int_{D}^{D_{\mathrm{max}}^{\prime}}n(D)\, f_{\mathrm{reach}}(D)\, f_{\mathrm{cross}}(D)\, dD.\label{eq:5}\end{equation}
This integral can be solved numerically taking into account that $D_{\mathrm{max}}^{\prime}=8$
km. The results are shown in Fig. \ref{fig:6} for $\bar{v}_{\mathrm{ej}}=200$
m/s and three different values of $\tau_{\mathrm{age}}$. The full
lines were obtained using the observed SFD, and the dashed lines using
the single power law model of the SFD. The true value of $N_{a>2.5}(D)$,
for a given $\tau_{\mathrm{age}}$, should be something between the
corresponding full and dashed lines. From Fig. \ref{fig:6}, it is
clear that the predicted $N_{a>2.5}$ for $D\gtrsim5$ km is at least
two orders of magnitude smaller that required to produce a body like
(21238). Therefore, we may conclude that, even if the crossing over
the J3/1 MMR is dynamically possible, it is highly improbable that
(21238) had reached its present orbit via this mechanism. On the other
hand, for $D\gtrsim3$ km the predicted $N_{a>2.5}$ is compatible
with the presence of (40521) in the middle belt, provided that $\tau_{\mathrm{age}}\gtrsim3.5$
Gy.

\section{Predicted vs. observed number of V-type asteroids in the middle belt\label{sec:4}}

Our results above have been compared to the observed population of
V-type asteroids in the middle belt, which is most likely incomplete.
Here we will estimate the debiased SFD of V-type asteroids in the
region $2.5<a\lesssim2.6$ AU, where (21238) and (40521) are found,
and will compare it to our estimates of $N_{a>2.5}$.

\begin{figure}[th]
\caption{The bias function (full line) in the region $2.5<a\lesssim2.6$ AU,
and its best fit (dashed line). The distributions $n_{\mathrm{know}}\Delta H$
(gray histogram) and $n_{\mathrm{sdss}}\Delta H$ (outlined histogram)
are shown for reference. Both distributions are normalized such that
$\sum n\,\Delta H=1$.}
\label{fig:7}\bigskip{}

\noindent \begin{centering}
\includegraphics[clip,width=1\columnwidth]{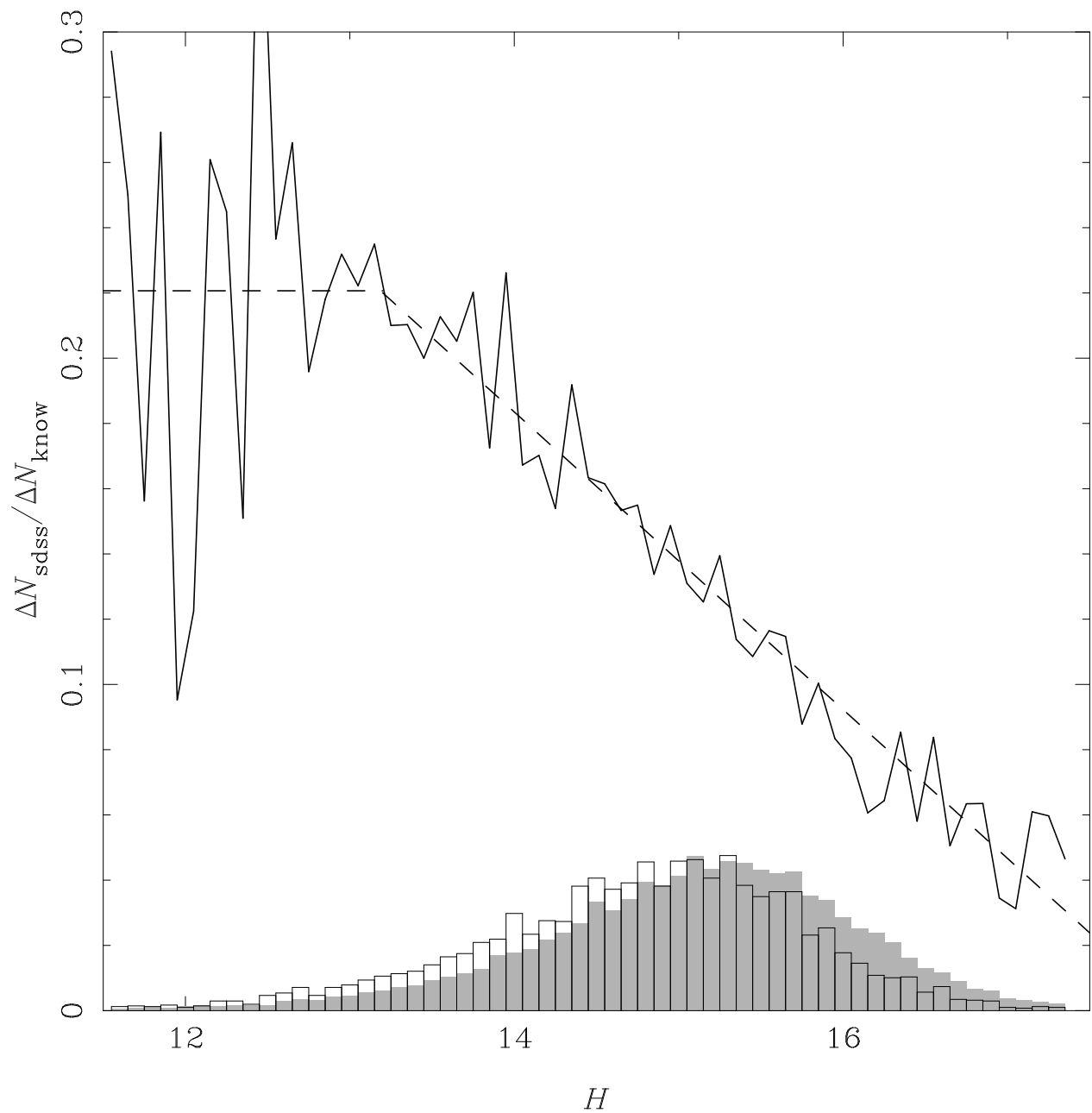}
\par\end{centering}
\end{figure}

In terms of absolute magnitude, the unbiased SFD of V-type asteroids
in the region of interest, $n_{\mathrm{V}}(H)\, dH$, is related to
the observed SFD, $n_{\mathrm{V}}^{obs}(H)\, dH$, through\begin{equation}
n_{\mathrm{V}}^{obs}(H)\, dH=b(H)\, n_{\mathrm{V}}(H)\, dH\label{eq:6}\end{equation}
where $b(H)$ is a bias function that we need to determine. This bias
function arises from two main effects: (i) the fact that the known
population of asteroids is complete only up to a certain size, and
(ii) the fact that the observed population $n_{\mathrm{V}}^{obs}$
is obtained from the SDSS that mapped only a fraction of the total
known population of asteroids. We may assume that effect (i) affects
the asteroid populations at both sides of the J3/1 MMR approximately
in the same way, so it can be ignored in our analysis. From effect
(ii), we have that

\begin{equation}
b(H)\simeq\frac{n_{\mathrm{SDSS}}(H)\, dH}{n_{\mathrm{know}}(H)\, dH}\label{eq:7}\end{equation}
where $n_{\mathrm{SDSS}}\, dH$ is the SFD of asteroids with $2.5<a\lesssim2.6$
AU contained in the 3rd. release of the SDSS-MOC, and $n_{\mathrm{know}}\, dH$
is the SFD of all the known asteroids in the same region, that can
be computed from the catalog of proper elements contemporary to the
SDSS-MOC. Figure \ref{fig:7} shows $b(H)$ (full line) and its best
fit (dashed line). The bias shows a linear dependence for $H\gtrsim13.0$,
and a constant value $\sim0.22$ for $H\lesssim13.0$. 

The observed SFD $n_{\mathrm{V}}^{obs}$ is poorly known since, according
to Table \ref{tab:1}, there are only two asteroids observed by the
SDSS-MOC in the range $2.5<a\lesssim2.6$ AU. Notwithstanding, we
can use additional observations from the SDSS-MOC to get a more populated
SFD. We follow the same approach as \citet{2006Icar..183..411R} to
identify V-type asteroids from the SDSS-MOC, but we disregard the
information in the $u$ band. This is justified since the $u$ band
is centered at $\simeq0.35\,\mu$m and it is not relevant for the
identification of V-type asteroids. Thus, we considered the SDSS-MOC
observations with errors less than 10\% only in the $g,r,i,z$ bands
and with any error in the $u$ band. With this approach, we find 27
V-type candidates in the middle belt, including the three asteroids
listed in Table \ref{tab:1}.

\begin{figure}[th]
\caption{The cumulative SFD of V-type asteroids with $2.5<a\lesssim2.6$ AU
observed by the SDSS (full black line) and the corresponding debiased
distribution (full gray line). The dashed line represents our predicted
$N_{a>2.5}$ for $\tau_{\mathrm{age}}=3.5$ Gy, computed from the
observed Vesta family SFD.}
\label{fig:8}\bigskip{}

\noindent \begin{centering}
\includegraphics[clip,width=1\columnwidth]{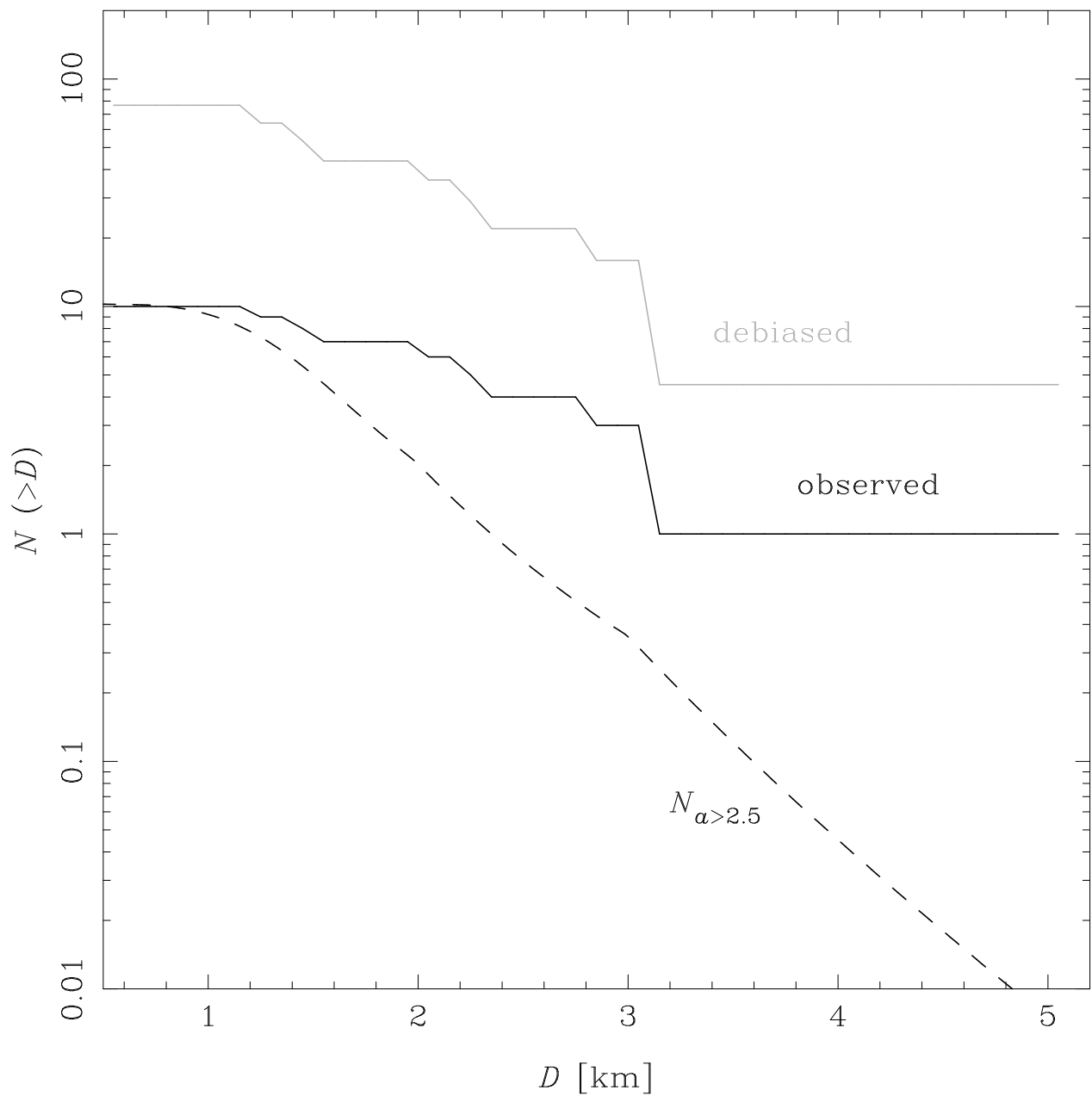}
\par\end{centering}
\end{figure}

The cumulative SFD of the 10 V-type candidates identified in the region
$2.5<a\lesssim2.6$ AU, and the corresponding debiased cumulative
SFD computed from Eqs. (\ref{eq:6}) and (\ref{eq:7}), are shown
in Fig. \ref{fig:8} (full lines). It is worth recalling that both
SFDs are affected by the completeness bias of the known asteroidal
population with $a>2.5$ AU. Therefore, we may compare these distributions
to our predictions of $N_{a>2.5}$ computed using the observed Vesta
family SFD which are affected by the same completeness bias. The comparison
to $N_{a>2.5}$ for $\tau_{\mathrm{age}}=3.5$ Gy (dashed line in
Fig. \ref{fig:8}) indicates that the predictions are $\gtrsim10$
times smaller than required in the studied size range. However, we
must bear in mind that the cumulative SFDs shown in Fig. \ref{fig:8}
correspond to asteroids that are candidate V-type according to the
SDSS colors. An unknown fraction of these bodies might not be true
V-types, but belong to other taxonomic classes like O-, Q-, R- or
S-type. Therefore, our predictions could actually account for $\gtrsim10$\%
of the total population of V-type asteroids in the middle belt, assuming
$\tau_{\mathrm{age}}\gtrsim3.5$ Gy.

\section{Conclusions\label{sec:5}}

In this paper, we presented spectroscopic observations in the visible
that confirm the existence of two V-type asteroids in the middle belt:
(21238) 1995WV7 and (40521) 1999RL95. We investigate whether these
two asteroids might have evolved from the Vesta family by slowly drifting
in semi-major axis due to the Yarkovsky effect and crossing over the
J3/1 mean motion resonance with Jupiter. Our results show that, in
spite of the remarkable instability of the J3/1 resonance, km-size
asteroids can cross it. The resonance crossing mechanism is probably
not sufficiently efficient to explain the presence of all km-size
V-type asteroids in the middle belt, but only some fraction ($\gtrsim10$\%)
of them. Most of these bodies either follow other paths from the Vesta
family or they come from a totally different source (e.g. \citealp{2007A&ASS..C}).
Notwithstanding, we cannot rule out the possibility that (21238) were
a rare exception: the only one 5-km V-type asteroid in the middle
belt that reached its present orbit by a very improbable, but not
impossible, crossing over the J3/1 resonance. Only the discovery of
more V-type asteroids in the middle belt, and the better knowledge
of the SFD of these bodies may shed some light on this problem.

\acknowledgements{This work has been supported through several grants and fellowships
by the Brazilian Council of Research (CNPq), the NASA's Planetary
Geology \& Geophysics Program, and the Rio de Janeiro State Science
Foundation (FAPERJ).}

\bibliographystyle{klunamed}
\bibliography{paper31}

\end{document}